\documentstyle[nato,psfig]{crckapb}

\begin{opening}
\title{The Effects of a Variable IMF on the Chemical Evolution  of the Galaxy} 

\author{C. CHIAPPINI$^1$}

\institute{Departamento de Astronomia - Observat\'orio Nacional\\
           R. Gal. Jos\'e Cristino 77 - Rio de Janeiro - 20921-400 - 
Brazil}

\author{F. MATTEUCCI$^2$}
\institute{Dipartimento di Astronomia - Universit\'a di Trieste\\
       Via G. B. Tiepolo 11 - 34131 - Trieste - Italy}

\end{opening}

\runningtitle{Effects of a variable IMF on the CE of the Galaxy}

\begin{document}

\begin{abstract}
In this work we explore the effects of adopting an initial mass 
function (IMF) variable in time on the chemical evolution of the Galaxy.
In order to do that we adopt a chemical evolution model which assumes
two main infall episodes for the formation of the Galaxy. 
We study the effects on such a model of different IMFs. First, we 
use a theoretical one based on the statistical 
description of the density field arising from random motions in the gas.
This IMF is a function of time as it depends on physical conditions of 
the site of star formation. We also investigate the behaviour of the 
model predictions using other variable IMFs, parameterized as a function
of metallicity.

Our results show that the theoretical IMF when applied to our model 
depends on time but such time variation is important only in the early 
phases of the Galactic evolution, when the IMF is biased towards 
massive stars. We also show that the use of an IMF which is a stronger 
function of time does not lead to a good agreement with the observational 
constraints suggesting that if the IMF varied this variation should have 
been small. Our main conclusion is that the G-dwarf metallicity distribution
is best explained by infall with a large timescale and a constant IMF, since
it is possible to find variable IMFs of the kind studied here, reproducing
the G-dwarf metallicity but this worsens the agreement with other
observational constraints.
\end{abstract}

\section{Introduction}

Observational constraints
are of fundamental importance to build a realistic chemical evolution model.
With respect to these constraints the last years have been of crucial 
importance (Pagel 1997) and, in the case of the Milky Way, 
the new observational data required a revision of the previous 
chemical evolution models (see Pagel and Tautvaisiene 1995 and 
Chiappini et al. 1997, hereinafter CMG, for a discussion on this point).

In particular, Gratton et al. (1996, 1999) showed that the 
distribution of the abundances of  $\alpha$-elements to Fe 
for a large homogeneous sample of stars in the solar neighbourhood 
seems to indicate a short timescale for the
evolution of the halo and thick disk phases and a sudden
decrease in the star formation in the epoch preceding the formation of the
thin disk. An analogous result was found by Fuhrmann (1998) 
for the [Mg/Fe] ratio. Moreover,
Beers and Sommer-Larsen (1995) have shown that the thick disk population 
extends to very low metallicities. 
Those are very important new information which stimulated us (CMG) 
to consider a different picture of Galaxy formation than those 
previously adopted.

Previous models, in fact, (e.g. Matteucci and Fran\c cois 1989)
were based on a pure colapse picture where the disk formed from gas
shed from the halo.
These models, however, are difficult to 
conciliate with the new results discussed above indicating a 
discontinuity between halo and thin disk. 
We than suggested the so-called {\it Two-Infall Model}, a model that 
assumes that the Galactic thin disk was not only formed from gas shed from 
the halo but was formed mainly from extra-galactic gas. 
We assume two main infall episodes, the first one is responsible for 
the formation of the population made of that fraction of the halo 
and thick disk stars which originated from a fast dissipative collapse,
such as suggested by Eggen et al. (1962).
The second infall episode forms the thin disk component with a 
timescale much longer than that of the halo formation. 

In this new picture the disk was 
formed slowly (with a timescale of 7-8 Gyrs at the solar vicinity) and from
inside-out (Matteucci and Fran\c cois 1989). 
A direct consequence of that is that at high
redshift we should expect to see smaller disks in size (Roche et al. 1998).
This long timescale for the formation of the thin disk at the 
solar vicinity, required to produce a good fit of the 
obseved G-dwarf metallicity distribution (Rocha-Pinto and Maciel 1996) was
also confirmed by recent chemical evolution models (eg. Portinari et al. 1998, Prantzos and Silk 1998, Chang et al. 1999) 
as well as by chemo-dynamical models
(eg. Hensler 1998) and is also in agreement with the results showed
in this conference by Carraro (2000).

The two-infall model adopted a constant
IMF. On the empirical grounds there is at present no clear direct 
evidence that the IMF in the Galaxy has varied with time.
A detailed discussion about possible observed variations
in the IMF in different environments is given by Scalo (1998), 
but such variations are comparable with the uncertainties
still involved in the IMF determinations. 
The present uncertainties in the observational results
prevent any conclusion against the idea of an universal IMF.

However, a variable IMF, which formed relatively more massive stars 
during the earlier phases of the evolution of the Galaxy compared to 
the one observed today in the solar vicinity, has often been suggested 
as being one of the possible solutions for the G-dwarf problem 
(namely the deficiency of metal-poor stars in the solar neighborhood
when compared with the number of such stars predicted by the simple model).
Such an IMF would also be physically plausible from the theoretical point
of view if the IMF depends on a mass scale such as the Jeans mass. 
Given the uncertainties in both theoretical and observational grounds,
the proposed IMFs can in principle be tested by means of a detailed 
chemical evolution model.

An example of a theoretical approach to the IMF problem is the one proposed
by Padoan et al. (1997 - hereinafter PNJ). Since random motions are 
probably ubiquitous in sites of star formation, PNJ suggested
to describe the formation of protostars as the 
gravitational collapse of Jeans masses in a density distribution 
shaped by random supersonic motions (but see Scalo et al. 1998).
This IMF was already tested in models of elliptical galaxies by
Chiosi et al. (1998) and they concluded that a strongly varying IMF
could be suitable for such galaxies.

Our goal was to address the question of what time-dependent 
IMF properties are allowed in order to still match the observational
constraints when adopting the two-infall model (CMG).
To do this we tested different IMFs in our chemical evolution code, going
from those where the variability is contained in the 
slope of the power-law, assumed to be a function
of the metallicity (parametrizations adopted by 
Scully et al. 1996, Matteucci and Tornamb\'e 1987), to the 
one by PNJ which predicts also a change in the stellar mass which
contributes most to the IMF as a function of time.

A detailed description of the {\it Two-Infall} model as well as of the
PNJ IMF and the hypothesis needed to introduce it in our chemical
evolution code are discussed in detail in Chiappini et al. 2000a 
(hereinafter CMP; see also CMG and PNJ).

\section{Results}

\subsection{The Solar Vicinity}

The two-infall chemical evolution model, published by
CMG (hereinafter Model A) has been modified to 
test the IMF proposed by PNJ (hereinafter Model B and C)
and the other two IMFs (Models D - Matteucci and 
Tornamb\'e 1987 and E - Scully et al. 1996). 
We address the reader to the CMP paper where a detailed description
of each one of the IMFs adopted here can be found.

The predictions of models A and B turn out to be very similar concerning
the observed properties of the solar vicinity (tables 2 and 3 of CMP). 
This is due to the fact that the PNJ IMF, 
when applied to our Galaxy, does not vary much over most of 
the Galactic lifetime which is a consequence of our assumptions 
of constant molecular cloud temperature during the Galactic 
evolution. In fact, if the cloud temperature and density have varied
strongly over the history of the Galaxy, the predicted IMF will also vary
more and this would certainly worsen the agreement with the observational
constraints considered here.

\begin{figure*}[h]
\centerline{\psfig{figure=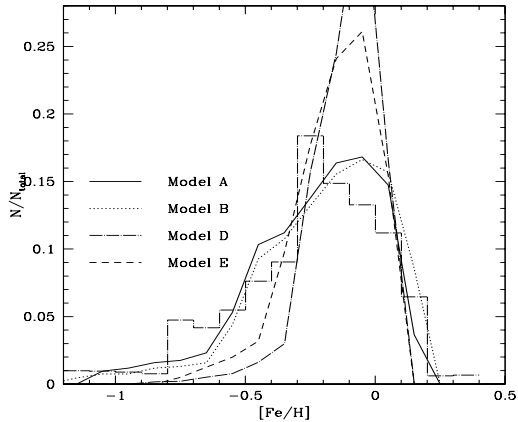,width=7cm,height=6cm}}
\caption{The G-dwarf metallicity distribution predicted
by models A, B, D and E. The data are
from Rocha-Pinto \& Maciel (1996)}
\end{figure*}

On the other hand, the other two IMFs (models D and E) vary substantially 
over the entire Galactic evolution. As a consequence of that, 
these IMFs do not give a good agreement with the solar vicinity 
observational constraints. Figure 1 shows the predicted G-dwarf 
metallicity distribution for models D and E compared 
with models A and B. Those models predict too few metal-poor stars 
and a higher solar metallicity peak than the observed one.
The predicted solar abundances are also not in agreement with 
the observed ones (Table 3 of CMP).

It is worth noting that here we adopted a long timescale for the 
formation of the solar vicinity (8 Gyrs) and that a better 
agreement with the G-dwarf observed metallicity distribution could 
in principle be achieved by adopting these variable IMFs in a closed
box model scenario. However, as recently showed by Martinelli and Matteucci
(1999), models which can fit the G-dwarf metallicity distribution
with a more variable IMF and which adopt a shorter infall timescale 
for the solar vicinity formation do not give good agreement with  
the other observed properties. Moreover, in a close box model the 
predicted gradients along the disk would be essentially flat.

From the comparison with the solar vicinity observational constraints we 
can confirm the result by Matteucci \& Tornamb\'e (1985) that only a 
constant IMF or an IMF that varied only at early times can be in 
agreement with the solar vicinity properties.

\subsection{The Galactic Disk}

The IMF in model B combined with an inside-out cenario for the
disk formation predicts a higher number of low mass-stars towards 
the galactic center, where the metallicity is higher (see Figure 8 of 
CMP). A direct consequence of this fact is the 
predicted flatter gradient with respect to the one of Model A. 
Figure 2a shows the oxygen abundance gradient as predicted by 
model A (CMG), model B (adopting the PNJ IMF) and model C 
(same as B but with star formation efficiency increasing with 
decreasing galactocentric distance). 
Model B predicts a flat gradient between 6 and 10 kpc,
and a small negative one in the inner part of the Galaxy ($ R < 6 kpc $). 
Model C, which adopts an increasing star formation efficiency $\nu$ 
towards the galactic center, predicts a steeper gradient inside the 
solar circle. In this case a the bimodal abundance gradient, 
steeper in the inner part and flatter outside, similar to the one 
found by CMG, is recovered. 
However, as it can be seen in Figure 2b, a model with a higher star formation
efficiency in the central parts (Model C) consumes the gas very fast thus
reaching the threshold density value very soon. As a consequence, model C 
predicts a flat gas density radial profile at variance with observations, 
whereas Model B predicts a gas density distribution very similar to the 
one predicted by model A and in better agreement with observations. 

\begin{figure*}[h]
\centerline{\psfig{figure=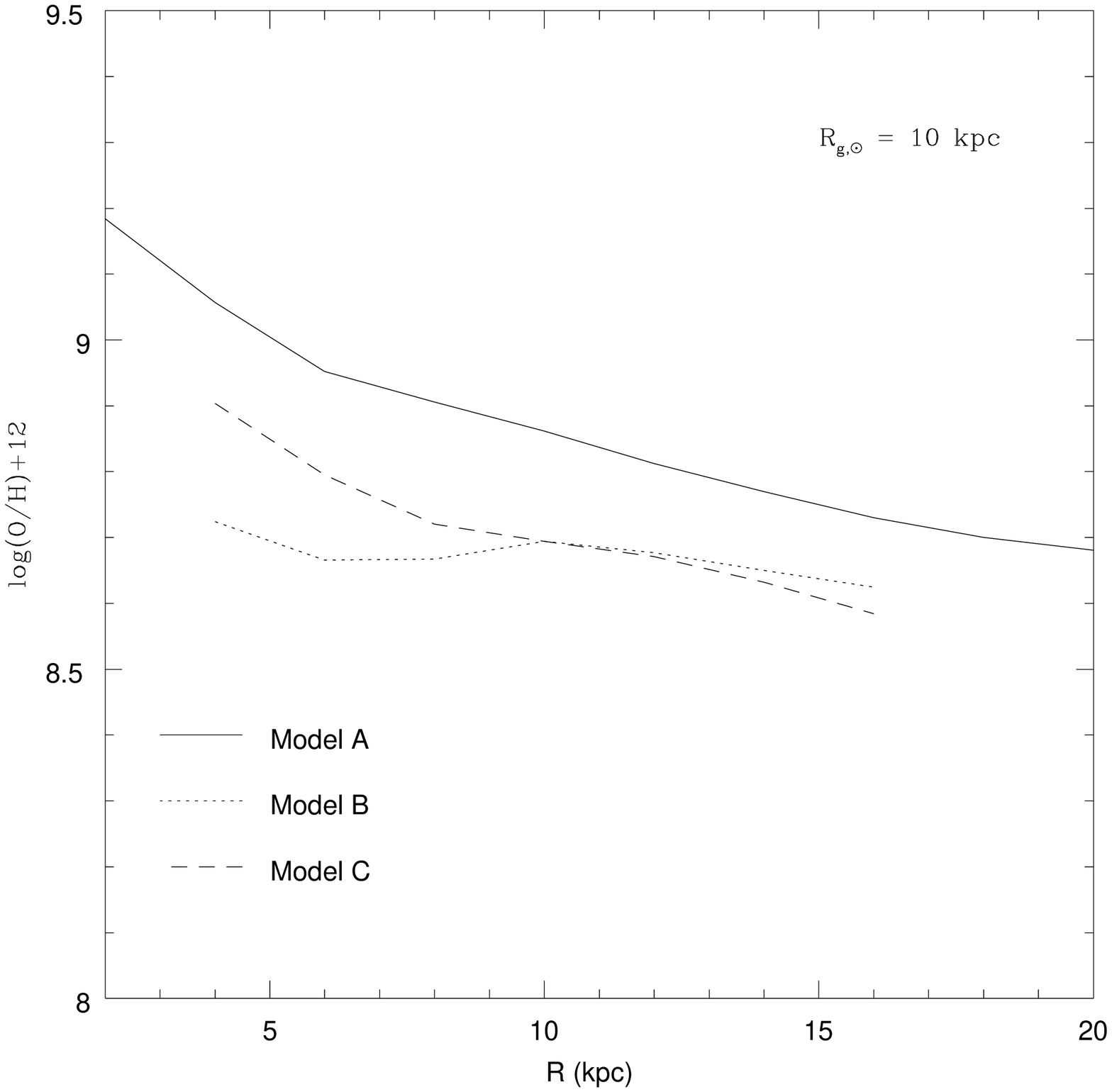,width=6.5cm,height=5cm}
\psfig{figure=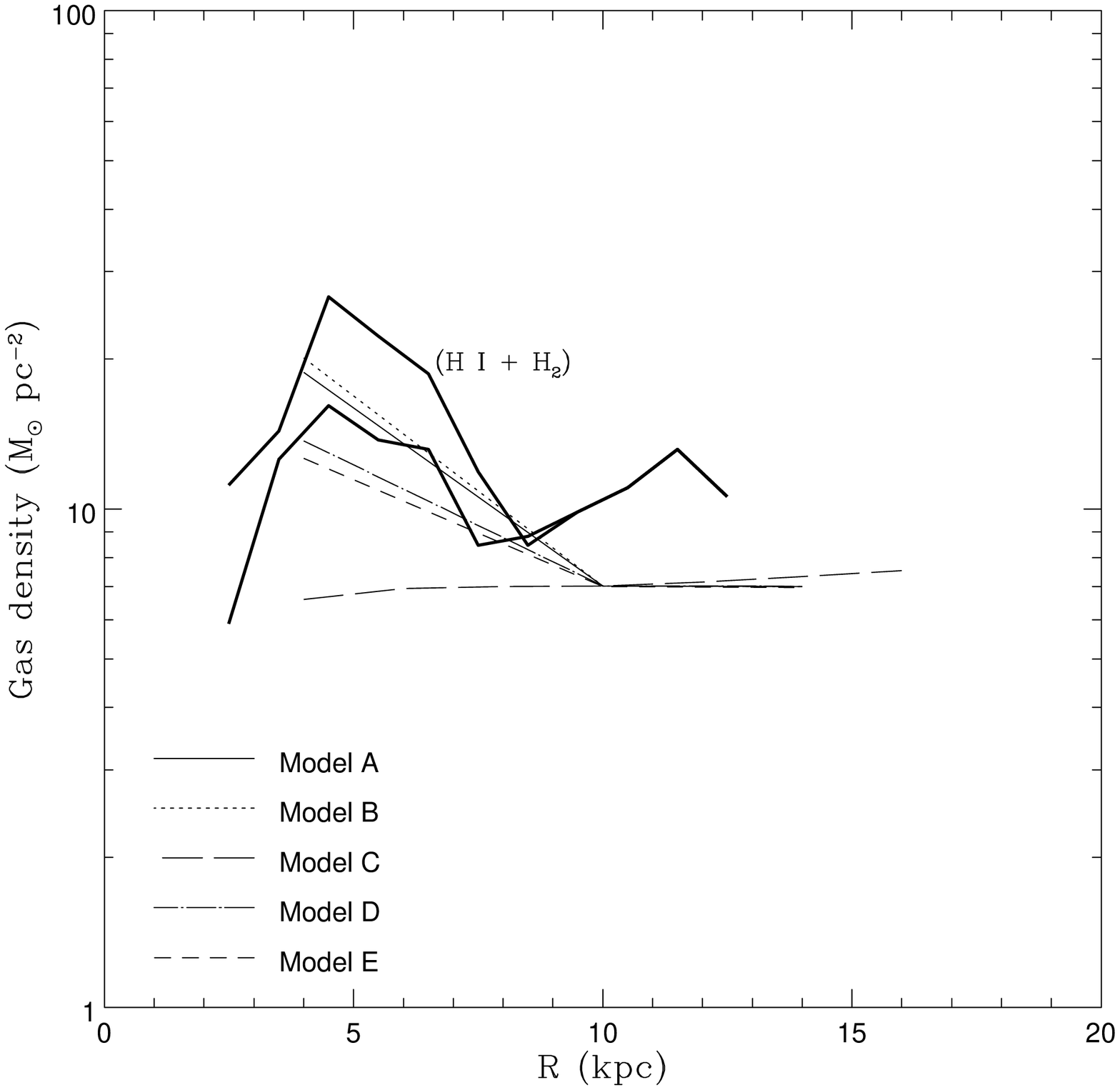,width=6.5cm,height=5cm}}
\caption{a) Oxygen abundance gradient predicted by models A, B and C;
b) Total surface gas density distribution given by Rana (1991) compared
with the one predicted by models A, B, C, D and E}
\end{figure*}

Figure 2b also shows models D and E predictions for the radial gas profile.
From this figure it can be seen that the only two models in good agreement
with the observed radial profile are models A and B. However, 
as shown before, model B predicts even flatter abundance gradients 
than model A. Models D and E predict a flatter radial gas profile 
than the observed one but in better agreement with the observations 
than the one predicted by model C. Moreover, the oxygen abundance 
gradient predicted by models D and E (see figure 11 of 
CMP) has a slope which is similar to the one 
obtained with model A. In any case, those two models do not give 
predictions in agreeement with the solar vicinity constraints as 
already mentioned in the previous section.

This is a very important result showing that only model A, 
with a constant IMF, is in agreement simultaneously with the 
solar vicinity and the disk observational 
constraints. The abundance gradients predicted by model A, 
although slightly flatter than the observed ones, are steeper in the inner 
parts of the Galactic disk and also steepen with time, 
in agreement with the recent results by Maciel \& Quireza (1999) (but
see Chiappini et al. 2000b for a detailed discussion on abundance 
gradients). 

\section{Conclusions}

The main results are summarized below:
\par\noindent
a) We tested the IMF proposed by PNJ in a model for the chemical 
evolution of the Milky Way (CMG) and we showed that this IMF
gives good agreement with the observed properties of the solar vicinity.
However, such an agreement is due to the fact that this IMF when applied
to the two-infall model shows a time variation that is important only 
in the early phases of Galactic evolution. This in turn is due to the 
simplifying assumptions adopted here, like neglecting the 
dependence of the IMF on the gas temperature which would produce 
more sharply varying IMF. In these early phases the IMF is biased 
towards massive stars.
\par\noindent
c) the PNJ IMF combined with the inside-out picture for the thin disk
formation predicts a gradient flatter than the one predicted by a 
model which adopts a constant IMF. 
This situation cannot be reversed by changing the 
SFR because in this case the abundance gradient is recovered but the gas 
density profile is destroyed. 
\par\noindent
d) Models which adopt IMFs strongly dependent on metallicity, 
thus simulating a dependence also on the gas temperature, are not
in agreement with the most important observational constraints of 
the solar vicinity and predict radial gas profiles at variance 
with observations, therefore they should be rejected.
\par\noindent
e) We conclude that a constant IMF and the assumption of a 
continuous infall onto the Galactic disk is still the best way 
to explain the observational constraints in the Milky Way 
including the G-dwarf metallicity distribution. A probable source
of the required infall could be the HVCs (see Burton and Braun 1999
for a discussion on this point).

{\small

\end{document}